\documentclass[aps,pra,twocolumn,groupedaddress]{revtex4-1}

\usepackage{graphicx}
\usepackage{latexsym}
\usepackage{bm,array}
\usepackage{amsfonts}
\usepackage{amssymb}
\usepackage{amsmath}
\usepackage{upgreek}
\usepackage{xcolor}
\usepackage{enumitem}

\newcommand{\mtx}[2]{\left(\begin{array}{#1}#2\end{array}\right)}

\begin{document}


\title{A classical formulation of quantum theory?}


\author{William F.~Braasch Jr.}
\affiliation{Department of Physics and Astronomy, Dartmouth College, Hanover, New Hampshire 03755, USA}

\author{William K.~Wootters}
\affiliation{Department of Physics, Williams College, Williamstown, Massachusetts 01267, USA}



\begin{abstract}
We explore a particular way of reformulating quantum theory in classical terms, starting with phase space rather than
Hilbert space, and with actual probability distributions rather than quasiprobabilities.  The classical picture we start with is epistemically restricted, in the spirit of a model introduced by Spekkens.  We obtain quantum 
theory only by combining a {\em collection} of restricted classical pictures.  Our main challenge in this
paper is to find a simple way of characterizing the allowed sets of classical pictures.  We present one promising approach to this problem and
show how it works out for the case of a single qubit.
\end{abstract}


\maketitle


\section{Introduction}

Much of Wojciech Zurek's research, including his research on quantum Darwinism, has been aimed at explaining the emergence of the classical world from
the quantum world.  This is of course an important endeavor, partly because, {as he has
pointed out,} quantum theory and classical physics seem almost incompatible at first sight.  
\begin{quote}
{``The quantum principle of superposition implies
that any combination of quantum states is also
a legal state.  This seems to be in conflict
with everyday reality: States we encounter
are localized.  Classical objects can be
either here or there, but never {\em both}
here and there.'' \cite{Zurek2009}}
\end{quote}
Indeed, it is an interesting fact that the standard formulation of quantum theory---with state vectors
in Hilbert space---looks as different as it does from the emergent 
classical picture. 
In this paper we take a step towards a reformulation of quantum theory
that looks more classical from the very beginning, being based on phase space rather than Hilbert space.  At the same time, we
wish to avoid the negative probabilities of the Wigner-function formulation,
which is the most common phase-space formulation of quantum theory.

We are motivated largely by the general observation that it is good to 
have alternative formulations of a well-established theory.  Alternative formulations can provide novel insights and new methods of analysis.  In the present case, we can also hope
that our classical-like formulation will ultimately provide another perspective on the quantum-to-classical transition.  

{Though we intend in future work to apply our 
methods to continuous quantum variables such as
position and momentum,} in this paper we restrict our attention to the case of systems normally described with a finite-dimensional Hilbert space. 
For us, this means that the phase spaces we use are {\em discrete}.  Specifically, we use the discrete phase space introduced in Ref.~\cite{Wootters1987}, which is simplest when the Hilbert-space dimension $d$ is 
prime.  In that case, the phase space is a $d \times d$ array of points, with axes---analogous to position and momentum
axes---labeled by elements of the field ${\mathbb Z}_d$,
{that is, the integers mod $d$}.  As we
explain in the following section, in this phase space it makes sense to speak of ``lines'' and ``parallel lines''.  Each line has exactly $d$ points, and there are $d+1$ ways of dividing the $d^2$ points of phase space into $d$ parallel lines.   

Our work is related to a construction due to Spekkens~\cite{Spekkens2007, Bartlett2012, Spekkens2016, Hausmann2021}.  Starting with the same discrete phase space, he defines an ``epistemically restricted classical theory'': the points of phase space are understood to be 
the actual, underlying states of the system, but an observer cannot know this state.  The most detailed description an observer can give is a uniform probability distribution over one of the lines.  Spekkens showed that many qualitative features of quantum theory can be captured by this model, but the model cannot fully imitate quantum theory because it is non-contextual.  

In a recent paper, we showed how one can construct a picture that borrows some of Spekkens' ideas but that accommodates the full quantum theory of a $d$-state system~\cite{Braasch2021}.  Specifically, we found that one can decompose the quantum description of a complete experiment---a preparation, a transformation (or a sequence of transformations), and a measurement---into 
a collection of classical descriptions, each
entailing certain epistemic restrictions
{similar to but subtly different from
the one imposed in Spekkens' model}.  There is one such classical description for each possible choice of what we call a ``framework.''  The framework defines the epistemic restrictions placed on the classical model.  Within each framework, we can 
imagine a classical observer whose picture of the experiment
is perfectly compatible with an ontological model in which
the system really does occupy a definite phase-space point 
at every moment, and in which a transformation is represented
by an ordinary set of transition probabilities in phase
space.  
Each classical
observer will compute their own prediction for the 
experiment, in the form of a probability assigned to
each possible outcome. We showed
how to combine these classical predictions to reconstruct the quantum prediction.  In a slogan, we say the quantum 
prediction is obtained by ``summing the nonrandom parts."
The meaning of this slogan will become clear in the following
section, but essentially, the nonrandom part of a probability value
is its deviation from the value one would
use under a condition of minimal knowledge.
(Thus the expression, ``the nonrandom part,''
is a kind of shorthand. We do not mean to imply that 
there is no element of randomness in
values of a probability that differ
from the
minimal-knowledge value.)  
{Intriguingly, the use of this unusual method of combining probability distributions allows us to reproduce the operational statistics of the non-commutative theory of quantum mechanics starting
with ordinary (commutative) classical probability theory.}

In Ref.~\cite{Braasch2021}, we were not able to come up with a simple set of criteria for determining precisely what sets of classical descriptions are allowed---we did specify a set of criteria, but it is not simple.  Such criteria are desirable if
our formulation of quantum 
theory is to be self-contained, that is, not dependent 
on concepts from the Hilbert-space formulation.  The primary aim of the present paper is to identify such a set of criteria.  

The rest of this paper is organized as follows.  In the next
section we review the formalism of Ref.~\cite{Braasch2021}: how the 
frameworks are defined, how one decomposes the quantum
description of an experiment into epistemically restricted
classical descriptions, and how the predictions based 
on these classical 
descriptions are combined to recover the quantum prediction.
In Section \ref{respectDeltas}, we write down four
equations showing how
any pair of components of an experiment---the components being
preparations, transformations, and measurement outcomes---can
be combined to
obtain either other components or an observable 
probability.  For example, a preparation followed
by a transformation constitutes another preparation.  
And for a preparation followed directly
by a yes-or-no test, there is an equation that yields the probability
of the outcome ``yes.''
The basis
of each of these four equations is the principle that the
nonrandom parts of the inputs should be summed to get the
nonrandom part of the output. At this point we ask our 
main question: to what extent do these four 
composition rules
determine the allowed sets of classical descriptions?
That is, to what extent do these equations characterize 
the structure of quantum theory for the $d$-state system? We find it useful to 
add a few auxiliary assumptions, but we
do not know whether all these assumptions
are necessary.  Conceivably a more
parsimonious set of postulates is possible.

In Section \ref{qubit},
we specialize to the case of a single qubit and ask whether the composition rules 
and auxiliary postulates of Section \ref{respectDeltas}
determine the quantum theory of this simple system.
We find that we recover either the standard
quantum theory for a qubit or a theory with
a discrete set of transformations.

Of course we would like to extend this approach to
all possible Hilbert-space dimensions and to composite
systems.  We discuss the possibilities for doing this
in the concluding section.

For the remainder of this Introduction, we review briefly
some of the earlier efforts to reconstruct quantum theory from basic principles, 
as well as other work on quantum theory in phase space 
and other
approaches to representing quantum theory in terms of
probability distributions.

Reconstructions of quantum theory can be traced back to Birkhoff and von Neumann~\cite{Redei1996}. In these initial forays, the focus was on mathematical axiomatizations~\cite{vonNeumann1932, Mackey1963, Piron1964, Ludwig1985}. However, it is appealing to think that quantum mechanics might be reconstructed by stipulating a set of principles in the spirit of Einstein's principles that lead to the theory of special relativity. This more
operationally oriented approach was ignited by Hardy~\cite{Hardy2001}.
In Hardy's axiomatization, the addition of the key word ``continuous'' to one of his principles differentiates quantum mechanics from classical probability theory.
Although the approach we describe here is different from Hardy's, that same key word rears its head as the distinguishing feature between quantum mechanics and a simpler theory, as we will see in Section \ref{qubit}.
Other important reconstruction efforts have likewise relied on operational or information theoretic principles~\cite{Clifton2003, DAriano2006, Goyal2008, Dakic2011, Masanes2011, Chiribella2011}. Recently, diagrammatic postulates have been used to reconstruct quantum theory~\cite{Selby2021}.

In another vein, attempts to pinpoint essential quantumness have taken the tack of augmenting classical physics with simple rules.
As we mentioned, Spekkens and collaborators have done this in a series of epistemically restricted classical theories used to support an epistemic interpretation of the quantum state~\cite{Spekkens2007, Bartlett2012, Spekkens2016}. 
Spekkens' model has previously been provided with a contextual extension but without fully capturing quantum theory~\cite{Larsson2011}. The model has also been shown to hold strong similarities to stabilizer physics~\cite{Catani2017, Catani2018}, thereby providing a link to a subtheory of quantum mechanics that plays an important role in quantum computing. Spekkens' model is naturally set in phase space, which 
provides the backbone for our work.

Quantum mechanics set in phase space has a history almost as long as quantum mechanics itself~\cite{Wigner1932, Moyal1947}. Again, the most commonly used phase-space representation
of a quantum state is the Wigner function.
{An interesting complementary strategy is to
invert the definition of the Wigner function to 
make classical mechanics look more like 
quantum theory \cite{Groenewold1946, Bracken2003}.} A number
of different discrete Wigner functions have been defined for finite-dimensional quantum systems~\cite{Wootters1987, Buot1974, Vourdas2017, Gibbons2004, Gross2006, Gross2007}.
In this paper, although our aim is to go beyond Wigner functions and use only non-negative probabilities, we do use concepts 
from the Wigner-function definition of 
Ref.~\cite{Wootters1987}.  

There are numerous reasons to study quantum mechanics in phase space.
In quantum optics, the appearance of negativity of the Wigner function
signals the onset of quantum behavior~\cite{Hudson1974, Kenfack2004}.
Negativity of the Wigner function has also been linked to contextuality, which is another famous notion of nonclassicality~\cite{Spekkens2008},
and to the power of quantum computing~\cite{Veitch2012,Bermejo2017,Delfosse2017}.

Tomography of quantum states is closely tied to the Wigner function.
We are effectively using tomographic representations of quantum states in this paper.
For this reason, our work is also closely related to the ``classical'' approach to quantum theory found in Refs.~\cite{Chernega2017, Mancini1996, Ibort2009}.
These authors have successfully described a large number of quantum phenomena from this perspective.
Our approach diverges from theirs in that we treat every aspect of a quantum experiment tomographically.


Certain subtheories of quantum mechanics have been proven to be nonnegative in the Wigner function representation. The stabilizer subtheory is one of them~\cite{Gross2006}, and sampling its nonnegative representation provides the basis for classical simulation of stabilizer physics~\cite{Veitch2012}.  Under certain assumptions, it has been
shown that the full quantum theory requires negativity in some aspect (preparation, transformation, or measurement) of a frame-theoretic generalization 
of the Wigner-function
representation~\cite{Ferrie2008, Ferrie2009, Ferrie2010}.
The Pusey-Barrett-Rudolph theorem is another expression of the limitation on representing quantum
mechanics with classical probability theory~\cite{Pusey2012}.
Nonetheless, in the search for new ways in which to simulate quantum systems, researchers have  found positive
probabilistic 
representations of quantum theory by loosening certain assumptions upon which the theorems are built~\cite{Lillystone2019, Raussendorf2020, Zurel2020, Cohendet1988, Cohendet1990, Hashimoto2007}. In another
setting, Fuchs and Schack have expressed quantum states
and transformations as probability distributions over the possible outcomes of a SIC-POVM~\cite{Fuchs2013}.

Again, our approach begins by decomposing the quantum
description of an experiment into a 
collection of classical
probabilistic descriptions, as we explain more fully
in the following section.

\section{A quantum experiment as a collection of classical experiments}
In this section we briefly review the formalism developed
in Ref.~\cite{Braasch2021}.  We begin with a bit of notation and terminology.

Let us assume for now that the system we are studying has
a Hilbert space with prime dimension $d$. 
We use Greek letters to label the points of 
the $d \times d$ phase space.
Each point $\alpha$ can be specified by its horizontal
and vertical coordinates, which we write as
$\alpha_q$ and $\alpha_p$, respectively,
to emphasize the analogy with position and momentum. 
Here $\alpha_q$ and $\alpha_p$ both take values in 
${\mathbb Z}_d$.
A {\em line} is the set of points $\alpha$ satisfying an equation
of the form $a \alpha_q + b \alpha_p = c$ for fixed
$a,b,c \in {\mathbb Z}_d$ with $a$ and $b$ not both zero. Two lines are parallel if they
can be specified by equations of this form differing only in 
the value of $c$. We refer to a line passing through the origin
as a {\em ray}.  

A point in phase space is not a valid quantum state,
but we find it extremely helpful to associate with each
point $\alpha$ a quasi-density matrix
$\hat{A}_\alpha$, which we call a
``phase point operator.''  This is a trace-one Hermitian matrix, but it
is not a legitimate density matrix because it can have
negative eigenvalues.  The matrices $\hat{A}_\alpha$ that we use
are the ones introduced in Ref.~\cite{Wootters1987} to define a discrete
Wigner function (see below).
The matrix $\hat{A}_\alpha$
for any odd prime $d$ is written as follows in terms of its components.
\begin{equation}
(\hat{A}_\alpha)_{kl} = \delta_{2 \alpha_q, k+l}\omega^{\alpha_p(k-l)},
\end{equation}
where $\omega = e^{2 \pi i/d}$ and the arithmetic in the
subscript of the Kronecker delta is mod $d$.  For $d=2$ there
is a special formula:
\begin{equation}
\hat{A}_\alpha = \tfrac{1}{2}\left[ \hat{I} + (-1)^{\alpha_p}\hat{X}
+ (-1)^{\alpha_q + \alpha_p}\hat{Y} + (-1)^{\alpha_q}\hat{Z} \right],
\end{equation}
where $\hat{X},\hat{Y},\hat{Z}$ are the Pauli matrices and $\hat{I}$ is the 
$2 \times 2$ identity matrix.
The $\hat{A}$ matrices are orthogonal in the Hilbert-Schmidt sense:
\begin{equation} \label{trAA}
\hbox{tr}(\hat{A}_\alpha \hat{A}_\beta) = d\delta_{\alpha\beta},
\end{equation}
and because there are $d^2$ of them, they serve as a basis
for the space of all $d \times d$ matrices.  In particular,
we can expand a density matrix $\hat{w}$ as a linear combination
of $\hat{A}$'s.  
\begin{equation}
\hat{w} = \sum_\alpha Q(\alpha | w) \hat{A}_\alpha.
\end{equation}
The coefficients $Q(\alpha|w)$ in this 
expansion constitute the {\em discrete Wigner function}
representing the given state.

The operators $\hat{A}_\alpha$ also have the following special property, which we
will use immediately 
in the following subsection. For any line $\ell$,
the average of the $\hat{A}$'s over that line is a one-dimensional
projection operator:
\begin{equation}
\frac{1}{d}\sum_{\alpha \in \ell} \hat{A}_\alpha =
|\psi_\ell\rangle\langle \psi_\ell |,
\end{equation}
where $|\psi_\ell\rangle$ is a state vector associated
with the line $\ell$. Since the $\hat{A}$'s are orthogonal to
each other, the $|\psi\rangle$'s associated with a complete
set of parallel lines---we call such a set a striation---constitute an
orthonormal basis for the Hilbert space.  Moreover,
because any two non-parallel lines intersect in exactly
one point, Eq.~(\ref{trAA}) guarantees that these bases
are mutually unbiased; that is, each basis vector is an 
equal-magnitude superposition of the vectors of any of 
the other bases.

\subsection{Defining the frameworks}
A ``framework'' is a mathematical structure
that determines what epistemic constraint
one of our classical probability distributions
must satisfy.  The introduction of the
concept of a framework is one way in which
our work differs from Spekkens' model.
In Spekkens' model, there is
just one classical world, and there is
one epistemic constraint that applies to
it.  In his model, for example, a
uniform probability distribution over any
line of the discrete phase space counts
as a legitimate epistemic state.
By contrast, what we are doing, roughly speaking, is to decompose this set of
possibilities into distinct cases---one for each possible slope of
a line---and to associate each of these
cases
with a different classical
world.  

We now explain specifically what kind of mathematical structure
constitutes a framework for each
component of an experiment---a preparation, a transformation,
or a measurement---and what epistemic restriction is
associated with each of these frameworks.

For either a preparation or a measurement, a framework is
simply a striation of the phase space---a 
complete set of parallel lines.  We label such a
striation with the symbol $B$, since each striation is
associated with an orthonormal basis.  For a given
framework $B$, the classical probability function 
representing a given preparation or measurement outcome
is required to be constant along each line of the
striation $B$---this is the epistemic restriction
associated with the framework $B$.  We define these 
restricted probability functions
in the next subsection.

To define the framework for a transformation, we need
to consider a special class of linear transformations
on the discrete phase space.  Let us think of a point
$\alpha$ as represented by a column vector with components
$\alpha_q$ and $\alpha_p$.  Then a linear transformation
is represented by a $2 \times 2$ matrix, with elements
in ${\mathbb Z}_d$, acting from the left on this column
vector.  A {\em symplectic} transformation is a linear
transformation that preserves the symplectic product
\begin{equation}
\langle \alpha, \beta \rangle 
= \alpha_p \beta_q - \alpha_q \beta_p.
\end{equation}
For the case we are considering, in which the phase space
has just two discrete dimensions, the symplectic transformations
are the same as the transformations whose matrices have
unit determinant.  


The number of symplectic matrices for any prime $d$ is
$d(d^2 - 1)$.  Our formulation is simplest if, among these
symplectic matrices, there exists a set $\mathcal{T}$ 
of just $d^2 - 1$ 
such matrices that 
has the ``nonsingular difference'' property: the difference between
any two matrices in $\mathcal{T}$ has nonzero determinant.  It turns out
that this condition allows for a 
particularly simple reconstruction
of a quantum transformation from the classical transition probabilities, defined
in the following section.  In Ref. \cite{Braasch2021},
the nonsingular difference is the only
property we require of the set $\mathcal{T}$, 
and, to our knowledge, it is not known whether such
a set of $d^2 - 1$ matrices exists for all prime $d$.  However, for our
present purposes, we also need $\mathcal{T}$ to constitute
a {\em group}, that is, a subgroup of the symplectic
group Sp(2,$d$).  It is known that
such a special subgroup exists for the values $d = 2, 3, 5, 7, 11$ but not for larger values~\cite{Chau2005}.
We can develop our formalism so as to apply to every 
prime dimension, regardless of whether there exists a special
set $\mathcal{T}$---indeed, we do this in 
Ref.~\cite{Braasch2021}.  (If no such set exists, we
use the full symplectic group and insert a factor of $1/d$ whenever we sum over the symplectic matrices.)  However, in the
present paper, for simplicity, we restrict
our attention to those dimensions 
for which such a special
subgroup exists.  This makes the equations simpler. 
In \hbox{Section
\ref{qubit}}, we specialize to the case of a single qubit,
for which we now write down explicitly the unique special
subgroup of symplectic matrices:
\begin{equation} \label{threelegal}
{\mathcal I} = \mtx{cc}{1 & 0 \\ 0 & 1}
\hspace{7mm} 
{\mathcal R} = \mtx{cc}{0 & 1 \\ 1 & 1}
\hspace{7mm}
{\mathcal L} = \mtx{cc}{1 & 1 \\ 1 & 0}
\end{equation}
One can verify that the difference between any
two of these matrices has nonzero determinant.
The choice of symbols comes from the fact that 
${\mathcal R}$ permutes the three nonzero points
by rotating them to the right, whereas ${\mathcal L}$
rotates them to the left.

Now, finally, we can say what we mean by a framework for
a transformation.  Again, let ${\mathcal T}$ be a group of 
$d^2 - 1$ symplectic transformations with
the nonsingular difference property.  
Then a framework for a transformation
is simply a symplectic matrix $S$ chosen from the set
${\mathcal T}$.  (When such a group does 
not exist, we let every symplectic matrix define
a framework.)

As we have said, for a preparation or a measurement outcome,
the associated probability function---defined in the following
subsection---will be required to be constant along each line
of the striation $B$ serving as the framework.  Something
similar happens for a transformation.  But instead of
working in phase space {\em per se}, we now imagine
ourselves working in the set of all {\em ontic transitions}
from one point to another.  Let us label such a 
transition as $\alpha \rightarrow \beta$.  There are
$d^4$ ontic transitions.  Now, just as a striation $B$
partitions the $d^2$ points of phase space into $d$
sets of $d$ points each, we can regard a symplectic
transformation $S$ as partitioning the $d^4$ ontic
transitions into $d^2$ sets, each comprising
$d^2$ transitions.  

Here is how this partitioning happens.  For a fixed
symplectic matrix $S$ and a given ontic transition
$\alpha \rightarrow \beta$, let the {\em displacement}
$\delta$ be defined by $\delta = \beta - S\alpha$. 
That is, $\delta$ is the extra displacement one needs
to get to $\beta$, once one has applied $S$ to $\alpha$.  
Keeping $S$ fixed, we define the ``displacement class'' 
associated with $\delta$ to be the set of all the ontic transitions
$\alpha \rightarrow \beta$ such that 
$\beta - S\alpha = \delta$.  For any given $S$, there are
$d^2$ displacement classes, each consisting of $d^2$
ontic transitions.  The framework $S$ entails the
following epistemic restriction: the classical transition probabilities characterizing
a given transformation must be {\em constant} within each
displacement class. In the following subsection, we
show how such a set of transition probabilities is to 
be defined.  

Consider now an entire experiment consisting of a preparation,
a transformation, and a measurement.  A framework for the
whole experiment is obtained by choosing a framework for
each component of the experiment.  We express a framework
${\mathcal F}$
for this experiment as the ordered triple $\mathcal{F} = (B', S, B)$, where
$B$ and $B'$ are the frameworks for the preparation and
measurement respectively.  (We read the ordered triple from
right to left, because in our equations, this is the order
in which the associated probability functions will appear.)

As it turns out, we will need to consider only a subset
of the possible combinations $(B', S, B)$, namely, those
for which the striation $B'$ is precisely the striation
obtained by applying $S$ to the striation $B$.  We call such
a combination a {\em coherent} framework for the 
experiment.  We {\em may} use other frameworks---ones that are
not coherent---but it turns out that such frameworks will
contribute nothing to our predictions for the outcome of
the experiment.  Similarly,
if the experiment includes two or more successive transformations, so that we
have a framework $(B', S_n, \ldots, S_1, B)$, we need to consider
only those frameworks for which $B' = S_n \ldots S_1B$.  

We are now ready to show how the quantum description of
a preparation, a transformation, or a measurement outcome
can be replaced by a set of epistemically restricted
classical descriptions.

\subsection{Decomposing the quantum description of an experiment into classical descriptions}
We begin with the case of a preparation.  The standard 
quantum description of a preparation is given by a 
density matrix $\hat{w}$.  We replace this single quantum
description with $d+1$ classical descriptions $R^B(\alpha | \hat{w})$,
one for
each striation $B$.  Each of these classical descriptions
is simply a probability distribution over phase space.  And
each of these probability distributions satisfies the
epistemic constraint associated with $B$: the distribution
must be constant along each line of $B$.  

The definition of $R^B(\alpha |\hat{w})$ in terms of the 
density matrix $\hat{w}$ is simple:
\begin{equation} \label{Rwdef}
R^B(\alpha | \hat{w}) = \frac{1}{d}\langle \psi_\ell | \hat{w} | \psi_\ell\rangle,
\end{equation}
where $\ell$ is the unique line in $B$ that contains
the point $\alpha$.  Again, $|\psi_\ell\rangle$ is the state
vector associated with the line $\ell$.  
It is not hard to show that $R$ is a properly normalized probability
distribution over phase space---that is, 
\begin{equation}
\sum_\alpha R^B(\alpha | \hat{w}) = 1,
\end{equation}
and it is clear that $R$ is constant over each line
in $B$.  It is possible to reconstruct $\hat{w}$ from 
the whole set of $R$'s, but in this paper our ultimate
aim is to work wholly with the classical descriptions.  
So we would like to think of the $R$'s as the primary
description of the preparation.  

We now move on to the case of a measurement (saving the
more complicated case of a transformation for later in this
subsection).  We are interested just in the probabilities
of the outcomes of a measurement, not in any change in the
system caused by the measurement.  A measurement in this sense
is represented in quantum theory by a POVM, that is, a set of
positive semidefinite operators on the $d$-dimensional
Hilbert space that sum to the identity.  Let $\hat{E}$ be
one element of such a POVM, corresponding to a particular
outcome of the measurement.  We now show how to replace
$\hat{E}$ with a set of classical probability functions
$R^B(\hat{E} | \alpha)$, one for each striation $B$.  In keeping
with the associated epistemic restriction,
the function $R^B(\hat{E} | \alpha)$ will
be constant along each line of $B$.  

The definition of $R^B(\hat{E} | \alpha)$ is similar to the one in Eq.~(\ref{Rwdef}).  
\begin{equation} \label{REdef}
R^B(\hat{E} | \alpha) = \langle \psi_\ell | \hat{E} | \psi_\ell \rangle,
\end{equation}
where $\ell$ is again the unique line in $B$ that contains
$\alpha$.  Informally, we think of $R^B(\hat{E} |\alpha)$
as the probability of the outcome $\hat{E}$ when the system
is at the point $\alpha$ (an illegal quantum state).  
Note that this function has a different normalization
from the classical probability distributions describing 
a preparation.  We can think of the uniform distribution
over phase space---with the value $1/d^2$ for each
point $\alpha$---as representing the completely mixed
state.  (This interpretation comes from the discrete Wigner function.)  So we expect the following
normalization:
\begin{equation}
\begin{split}
&\sum_\alpha \left[ R^B(\hat{E} | \alpha) \times \frac{1}{d^2}\right] = 
\hbox{tr}\left[ \hat{E} (\hat{I}/d) \right] = \frac{1}{d}\hbox{tr}\hat{E}\\
\implies &\sum_\alpha R^B(\hat{E} | \alpha) = d\,\hbox{tr}\hat{E}.
\end{split}
\end{equation}
And one can see from Eq.~(\ref{REdef}) that the function is indeed normalized
in this way.  

We now turn our attention to the case of a transformation.  
In general, the quantum description of a normalization-preserving
transformation is given by a completely positive, 
trace-preserving map, which in turn can be specified by
a set of Kraus operators.  We will replace this description
by a set of classical probability distributions.  We
restrict our attention to operations that preserve the
Hilbert-space dimension, and we restrict our attention
to {\em unital} transformations, that is, transformations
that leave the completely mixed state unchanged.

We begin by defining a set of {\em transition quasiprobabilities} that characterize a given transformation.  For an 
operation ${\mathcal E}$, these are defined by
\begin{equation}\label{Qdef}
Q_{{\mathcal E}}(\beta|\alpha) = \frac{1}{d}\hbox{tr}
\left[ \hat{A}_\beta \mathcal{E}(\hat{A}_\alpha)\right].
\end{equation}
In particular, if ${\mathcal E}$ 
is a unitary transformation, we have
\begin{equation}\label{QUdef}
Q_{\mathcal E}(\beta|\alpha) = \frac{1}{d}\hbox{tr}
\left[ \hat{A}_\beta \hat{U} \hat{A}_\alpha \hat{U}^\dag \right].
\end{equation}
In a discrete-Wigner-function formulation, we can 
interpret $Q_{\mathcal E}(\beta|\alpha)$ as the quasiprobability
that a system at the point $\alpha$ will move to the
point $\beta$ when the transformation $\mathcal{E}$ is applied.  Thus, 
if the transformation is applied to a 
system described by the Wigner function
$Q(\alpha|\hat{w})$, the resulting Wigner function
$Q(\beta|\mathcal{E}(\hat{w}))$ is given by
\begin{equation}  \label{Wignertrans}
Q(\beta|\mathcal{E}(\hat{w}))
= \sum_\alpha Q_{\mathcal E}(\beta|\alpha)
Q(\alpha|\hat{w}).
\end{equation}
Though $Q_{\mathcal E}(\beta|\alpha)$ plays the role of a probability in this equation, it is not a probability since it can take
negative values~\cite{Braasch2020}.  It is, however, normalized like a 
probability distribution: $\sum_\beta Q_{\mathcal E}(\beta | \alpha) =1$.  
Because our transformations are unital, $Q_{\mathcal E}$ is also 
normalized over its second argument: 
$\sum_\alpha Q_{\mathcal E}(\beta | \alpha) = 1$.

We use $Q_{\mathcal E}(\beta|\alpha)$ to define our classical transition probabilities
(which are indeed non-negative).  Again, the framework for
a transformation is specified by a symplectic transformation $S$
chosen from the set ${\mathcal T}$ defined above.  In the
framework $S$, the probability that a system at point $\alpha$
will move to $\beta$ is given by
\begin{equation}\label{QtoR}
R_{\mathcal E}^S(\beta | \alpha) = \frac{1}{d^2} 
\sum_\mu Q_{\mathcal E}(S\mu + \delta | \mu),
\end{equation}
where $\delta = \beta - S\alpha$.  That is, we get
$R_{\mathcal E}^S(\beta | \alpha)$ simply by averaging $Q_{\mathcal E}(\beta | \alpha)$
over the displacement class $\delta$ in which the
ontic transition $\alpha \rightarrow \beta$ lies.  
By definition, then, $R_{\mathcal E}^S(\beta | \alpha)$ is constant
over each displacement class.  

What is much less obvious is that $R_{\mathcal E}^S(\beta | \alpha)$
is always non-negative.  This is proved in Ref.~\cite{Braasch2021} and we
do not repeat the proof here.  (For the
special case $d=2$, the non-negativity
{\em depends} on using the special
subgroup $\mathcal{T}$ of symplectic
matrices.  For odd primes, $R^S_{\mathcal E}$
is non-negative for any symplectic $S$.)  We also show in that paper
how to reconstruct the quantum operation $\mathcal{E}$ from
the entire set of $R_{\mathcal E}^S$'s.  

We now have all the ingredients we need for a classical
description of a whole experiment, within a specified framework.
Let us suppose the experiment consists of a preparation, 
followed by a transformation, followed by a measurement.
In terms of standard quantum mechanical concepts, we can compute the probability of a 
particular outcome via the equation
\begin{equation}
P(\hat{E}|\mathcal{E},\hat{w}) = \hbox{tr}\left[ \hat{E} \mathcal{E}(\hat{w}) \right],
\end{equation}
where $\hat{w}$ is the initial density matrix, $\mathcal{E}$ is the transformation, and $\hat{E}$ is the POVM element representing
the outcome.  

Within the classical framework $(B', S, B)$, we can 
try to compute the same probability by writing
\begin{equation}  \label{classical1}
P(\hat{E} | \mathcal{E}, \hat{w}) \stackrel{?}{=} \sum_{\alpha\beta} R^{B'}(\hat{E}|\beta)
R_{{\mathcal E}}^S(\beta | \alpha) R^B(\alpha | \hat{w}).
\end{equation}
Note that we are combining the probabilities in the
standard way.
And again, every function inside the sum is non-negative and
properly normalized, so the resulting probability is at least
a legitimate probability.  But it is not the correct value.
This is largely because the classical story associated
with a specific framework is by no means the whole 
story.  We need the predictions obtained from {\em all} 
the coherent frameworks in order to recover the quantum 
prediction.  We show how this is done in the following 
subsection.

\subsection{Recovering the quantum prediction: summing the nonrandom parts}
The formula for reconstructing the quantum prediction from
the whole set of classical predictions is quite 
simple.  As we noted in the Introduction, it depends on the concept of the ``nonrandom 
part'' of a probability, which we now explain.

For a probability distribution $R(\alpha)$ over the discrete
phase space, we define the nonrandom part 
$\Delta R(\alpha)$ to be
the deviation from the uniform distribution:
\begin{equation}
\Delta R(\alpha) = R(\alpha) - \frac{1}{d^2}.
\end{equation}
For the probability of a measurement outcome $\hat{E}$,
we define the nonrandom part by subtracting off
the probability we would assign to the
outcome $\hat{E}$ if we were starting with
the completely mixed state, or in phase-space
language, if we were starting from the 
uniform distribution over phase space.
Thus we have
\begin{equation}
\Delta R^B(\hat{E}|\alpha) = R^B(\hat{E}|\alpha) - \frac{1}{d^2}
\sum_\gamma R^B(\hat{E}|\gamma),
\end{equation}
or, for an expression using standard quantum mechanical terms,
\begin{equation}
\Delta P(\hat{E}|\hat{w}) = P(\hat{E}|\hat{w}) - \frac{1}{d}\hbox{tr}\hat{E}.
\end{equation}
For all these cases,``$\Delta$'' means that 
we are subtracting the ``random part'' of
the given probability, that is, the
value we would assign to the probability
under a condition of minimal knowledge.

Let us now consider an experiment consisting of a
preparation $\hat{w}$, a transformation $\mathcal{E}$, and a measurement,
one of whose possible outcomes is $\hat{E}$.
We showed in Ref.~\cite{Braasch2021} that
we recover the quantum mechanically predicted probability
of the outcome $\hat{E}$ via the following formula:
\begin{equation} \label{sumnonrandom}
\Delta P(\hat{E}|\mathcal{E},\hat{w}) =
\sum_{\mathcal F} \Delta P^\mathcal{F}(\hat{E}|\mathcal{E},\hat{w}),
\end{equation}
where the sum is over all coherent frameworks
${\mathcal F} = (B', S,B)$, and
\begin{equation}  \label{classical2}
P^\mathcal{F}(\hat{E}|\mathcal{E},\hat{w})
= \sum_{\alpha\beta}
R^{B'}(\hat{E}|\beta)R^S_{{\mathcal E}} (\beta|\alpha)
R^B(\alpha | \hat{w}).
\end{equation}
That is, within each framework, we compute the probability
of $\hat{E}$ in an utterly standard way.  What is nonstandard
is that we then combine these various classical predictions
by summing the nonrandom parts. 

One component of the derivation of 
Eq.~(\ref{sumnonrandom}) is the formula
that inverts Eq.~(\ref{QtoR}), which is
also proved in Ref.~\cite{Braasch2021}:
\begin{equation} \label{RtoQ}
\Delta Q_\mathcal{E}(\beta|\alpha)
= \sum_S \Delta R_\mathcal{E}^S(\beta|\alpha).
\end{equation}
We will find this equation useful in 
Section 4 below.

\section{The composition rules and their
role as foundational postulates} \label{respectDeltas}
In the preceding section, we started with the standard
quantum mechanical description of each component of an
experiment and then defined our classical probability
functions in terms of the associated quantum concepts.
Our ultimate aim, though, is to develop a self-contained
formulation of quantum theory, in which the basic 
objects are epistemically restricted classical probability
functions.  This means that we cannot rely on the 
standard concepts of quantum theory to determine which
sets of classical probability functions are allowed. 
We must find criteria that are independent of
the vectors and operators of Hilbert space. 
It is to this aim that we now turn our 
attention.  In this section,
therefore, we switch to an operational understanding
of the symbols $w$, $\mathcal{E}$ and $E$.
We use those
symbols to refer to a preparation, a transformation, and
a measurement outcome---processes
and events one can observe in a lab---and not
to any particular
mathematical objects. The absence of hats on the
symbols is a notational indication of this switch.

Again, the calculation in Eq.~(\ref{classical2}), in which
we compute
the probability of the outcome $E$ from the perspective of
one of our classical observers, is quite ordinary---all
the probabilities are being used in the standard way.
It is only when we combine the classical predictions, via
Eq.~(\ref{sumnonrandom}), that we combine
probabilities in a way that we would never 
do classically---by summing the nonrandom parts.  
We are inclined, then, to regard the summing of the nonrandom
parts as the essentially quantum mechanical component
of our formulation.  We do not claim to fully understand
the significance of this procedure.  But it does seem to
capture what is quantum mechanical about our formalism,
just as the superposition principle can be understood
as the quintessential quantum mechanical feature of the 
usual formulation. 
{We do not mean to imply that our rule is in any way an 
expression of the superposition principle but only that 
we are giving our rule a fundamental status in the 
mathematical formalism.}
(The superposition principle is
quite foreign to our approach, since we are working
only with probabilities and not with amplitudes.)

This circumstance leads us to ask whether the procedure of summing nonrandom parts can be used as a foundational 
principle, which could determine
which sets of probability distributions are permitted.

To that end, let us consider the following four equations---the ``composition rules''---all of which follow from the definitions of the preceding 
section but all of which also make sense without
reference to any
Hilbert-space concepts.
In these equations, the symbol $\Delta$ is 
consistently used to indicate the nonrandom
part of whatever follows it. 
\begin{enumerate}
\item Combining a preparation with a transformation to 
obtain another preparation.
\begin{equation} \label{respect1}
\begin{split}
\Delta & R^{B'}(\beta|\mathcal{E}(w)) =  \\ &\sum_{\{(S,B)|SB=B'\}}\Delta \left[ \sum_\alpha R_{\mathcal E}^S(\beta|\alpha)R^B(\alpha|w)\right].
\end{split}
\end{equation}
\item Combining two transformations in sequence to obtain
another transformation.
\begin{equation} \label{respect2}
\begin{split}
&\Delta  R^S_{\mathcal{E}_2 \circ \mathcal{E}_1}(\gamma|\alpha)
=\\
&\sum_{ \{(S_1,S_2) | S_1S_2 = S \}  }   \Delta \left[ \sum_\beta R_{\mathcal{E}_2}^{S_2}(\gamma|\beta)R_{\mathcal{E}_1}^{S_1}(\beta|\alpha)\right].
\end{split}
\end{equation}
\item Combining a transformation with a measurement outcome
to obtain another measurement outcome.
\begin{equation} \label{respect3}
\begin{split}
&\Delta R^{B'}(E'|\alpha)= \\
&\sum_{\{(B,S)|S^{-1}B=B'\}}\Delta \left[ \sum_\beta R^B(E|\beta)R_{\mathcal E}^S(\beta|\alpha) \right],
\end{split}
\end{equation}
where $E'$ is the measurement outcome that is equivalent to applying $\mathcal{E}$
and then getting the outcome $E$.
\item Combining a preparation $w$ with a measurement outcome $E$
to obtain the probability $P(E|w)$ of the outcome $E$ given
the preparation $w$.
\begin{equation}  \label{respect4}
\Delta P(E|w) = \sum_B \Delta \left[ \sum_\alpha R^B(E|\alpha)R^B(\alpha|w) \right].
\end{equation}
\end{enumerate}
We can summarize all of these equations by saying that
in any combination of components of an experiment, one
always sums the nonrandom parts of the classically
expected results, the sum being over all frameworks
that are consistent with the framework of the 
resulting classical probability function (or, in the
last case, all frameworks that are coherent).  

The equations listed above are all correct as statements
within quantum theory, but our 
question now is whether they are {\em sufficient} to
pick out the allowed sets of $R$ functions.  

They may not be fully sufficient.  These
equations set conditions on the whole
system of probability distributions,
describing
all the 
components of an experiment, and
there could be trade-offs among these
components.  It is conceivable, for example, that by being
more restrictive in what we allow for measurements, we
can be more generous in what we allow for preparations.  
In this paper, we avoid some of this worry by making the following
three auxiliary assumptions, but we are not 
certain whether all these auxiliary assumptions are 
necessary.  
\begin{enumerate}[label=\Alph*.]
\item For each preparation $w$ with probability functions $R^B(\alpha | w)$, there is a corresponding measurement
outcome $E$ with probability functions 
$R^B(E | \alpha) = dR^B(\alpha | w)$, and the random part
of $R^B(E | \alpha)$ is $1/d$. 
\item For an invertible transformation ${\mathcal E}$
with probability functions $R^S_{\mathcal E}(\beta | \alpha)$,
the probability functions of the inverse are
$R^S_{\mathcal{E}^{-1}}(\beta | \alpha)
= R^{S^{-1}}_{\mathcal E}(\alpha | \beta)$.
\item Every preparation consistent with Eq.~(\ref{respect4})
and Assumption A is physically possible.  Moreover, the complete system of preparations,
transformations, and measurement outcomes must be 
{\em maximal}.  That is, it should not be possible to
add any other transformation or measurement outcome
without violating Eqs.~(\ref{respect1}--\ref{respect4})
or one of our Assumptions.  
\end{enumerate}
Assumption A minimizes the likelihood of precisely the kind of trade-off
we described above.  Assumption B gives the most natural 
definition of the inverse in our
formalism.  The spirit behind Assumption C is that we
are starting with a picture in which all properly normalized
probability functions are allowed.  The
composition rules and the auxiliary
assumptions are intended simply to 
restrict the set of such functions, and we do not want
to restrict it more than necessary.

To see how a proposed set of $R$ functions might run
afoul of the composition rules and the
auxiliary assumptions, suppose that for a single
qubit, we were to say that there exists a preparation $w$
such that for each striation $B$,
\begin{equation}
R^B(\alpha|w) = \left\{ \begin{array}{l} \tfrac{1}{2} \; \hbox{if $\alpha$ lies on the ray in $B$}\\ 0 \; \hbox{otherwise}\end{array} \right.
\end{equation}
Then by Assumption A, there exists a measurement outcome $E$ such that
\begin{equation}
R^B(E|\alpha)= \left\{ \begin{array}{l} 1 \; \hbox{if $\alpha$ lies on the ray in $B$}\\ 0 \; \hbox{otherwise}\end{array} \right.
\end{equation}
These functions are perfectly consistent with 
the epistemic constraint---each is 
constant along each line of its striation $B$---but they are 
not consistent with Eq.~(\ref{respect4}): the
computed probability for the outcome $E$, given
the preparation $w$, comes out to be 2, which
is not a legitimate value for a probability.  
Our question is whether similar considerations
will rule out all other sets of $R$'s that do
not correspond to legitimate quantum states and
processes.  

We have by no means answered this question in 
general, but we do have some answers for the 
special case of a single qubit.  They are the 
subject of the following section.

\section{The case of a single qubit} \label{qubit}
Here we show how Eqs.~(\ref{respect1}--\ref{respect4}) and Assumptions
A, B, and C apply to the case $d=2$.  

\subsection{The allowed preparations and measurements}
For now let us continue to take $d$ to be any prime number.
Let the functions $R^B(\alpha | w)$ describe a preparation $w$.
Then by Assumption A, there is a measurement outcome $E$
described by $R^B(E | \alpha) = dR^B(\alpha | w)$.  
Inserting this $w$ and $E$ into Eq.~(\ref{respect4}),
we get
\begin{equation}
\Delta P(E|w) = \sum_B \Delta\left[ \sum_\alpha R^B(E|\alpha)R^B(\alpha|w)\right].
\end{equation}
Each term in this equation that is preceded
by $\Delta$ is a probability assigned
to the outcome $E$.
So the $\Delta$ tells us to subtract $1/d$ (as is also
specified in Assumption A).  Collecting the
constant terms on the right-hand side, we have
\begin{equation}
P(E|w) = \sum_{B,\alpha}
\left[R^B(E|\alpha)R^B(\alpha|w)\right] -1.
\end{equation}
Now we replace $R^B(E|\alpha)$ with $dR^B(\alpha | w)$ to get
\begin{equation}
P(E|w) = d \sum_{B,\alpha} 
 R^B(\alpha | w)^2  - 1.
\end{equation}
In order to prevent $P(E|w)$ from being larger than 1,
we need to insist that
\begin{equation} \label{overd}
\sum_{B,\alpha} R^B(\alpha|w)^2 \le \frac{2}{d}.
\end{equation}
This condition must hold for every prime $d$.
As we now show, for the case of a single qubit,
it completely defines the set of allowed preparations.

For a qubit, Eq.~(\ref{overd}) becomes simply
\begin{equation} \label{qubit2overd}
\sum_{B,\alpha} R^B(\alpha | w)^2 \le 1.
\end{equation}
Suppose we have a set of functions $R^B(\alpha | w)$
satisfying this inequality.  Let us define the
quantities $r_x$, $r_y$, and $r_z$ as follows.
\begin{equation}
\begin{split}
&r_x = \sum_\alpha (-1)^{\alpha_p} R^X(\alpha | w) \\
&r_y = \sum_\alpha (-1)^{\alpha_q + \alpha_p} R^Y(\alpha|w) \\
&r_z = \sum_\alpha (-1)^{\alpha_q} R^Z(\alpha | w), 
\end{split}
\end{equation}
where $X$, $Y$, and $Z$ are the horizontal, diagonal,
and vertical striations, respectively.
These equations can be inverted to give
\begin{equation} \label{Rfromr}
\begin{split}
&R^X(\alpha|w) = \tfrac{1}{4}\left[ 1 + (-1)^{\alpha_p} r_x \right] \\
&R^Y(\alpha|w) = \tfrac{1}{4}\left[ 1 + (-1)^{\alpha_q + \alpha_p} r_y \right] \\
&R^Z(\alpha|w) = \tfrac{1}{4}\left[ 1 + (-1)^{\alpha_q} r_z \right].
\end{split}
\end{equation}
One can see from Eq.~(\ref{Rfromr}) that
\begin{equation}
\sum_{B,\alpha} R^B(\alpha|w)^2 = 
\tfrac{3}{4} + \tfrac{1}{4}\left(r_x^2 + r_y^2 + r_z^2\right).
\end{equation}
So Eq.~(\ref{qubit2overd}) is equivalent to the condition
that the vector $\vec{r} = (r_x, r_y, r_z)$ has length
no greater than 1. 

From the definition of $R^B(\alpha | w)$ in Section 2,
one can show that the $R$'s given in \hbox{Eq.~(\ref{Rfromr})}
correspond to the density
matrix 
$\hat{w} = \tfrac{1}{2}(I + \vec{r}\cdot\hat{\vec{\sigma}})$,
where $\hat{\vec{\sigma}}$ is the vector of Pauli matrices.  
We see, then, that the condition (\ref{qubit2overd}) 
is indeed sufficient to restrict the set of $R$'s to 
their proper range (that is, to the range
$|\vec{r}| \le 1$).  
Assumption C then tells us that the entire set of such preparations
is allowed.  In this way we recover the Bloch sphere.  


Do our assumptions also pick out the valid
measurement outcomes?  In the standard quantum
formalism, we can characterize the allowed 
POVM elements $\hat{E}$ by the following condition:
an operator $\hat{E}$ is a valid POVM element if and 
only if the quantity 
\begin{equation} \label{PEwq}
P(E|w) = \hbox{tr}(\hat{E} \hat{w})
\end{equation}
lies in the
interval $[0,1]$ for every density matrix
$\hat{w}$. Now, Eq.~(\ref{respect4}) is simply an
expression of Eq.~(\ref{PEwq}) in our formalism.
So the condition that the $P(E|w)$ appearing
in \hbox{Eq.~(\ref{respect4})} must be in the range
$[0,1]$ is equivalent to the quantum condition
we have just stated.
\hbox{Eq.~(\ref{respect4})} thus picks out the valid
measurement outcomes, as long as we know what
the valid preparations are.  And this we do know,
as we have seen in the preceding paragraph.

\subsection{The allowed invertible transformations}
Here we aim to determine what set or
sets of invertible
transformations on a qubit are consistent
with our assumptions.

We begin by noting that for any invertible
transformation $\mathcal{E}$, we can derive from 
Assumption B that $R_{\mathcal E}^S(\beta|\alpha)$ is normalized over its
second index as well as its first. 
(As we have noted earlier, the same is true for any unital transformation, a concept that still makes
sense in our phase-space setting.) We 
use this fact a few times in what follows.

Our next step is to derive the representation of the identity transformation $R^{S}_I(\beta|\alpha)$. 
For $d=2$ there is a valid preparation given by the following probability distributions:
\begin{equation}
\begin{split}
R^X(\alpha|{w})  = R^Y(\alpha|{w}) =& \frac{1}{4} \;
\bgroup 
\def\arraystretch{1.7} 
\begin{tabular}{|c|c|}
\hline
$\;\; 1 \;\;$ & $\;\; 1 \;\;$ \\
\hline
$1$ & $\;\; 1 \;\;$   \\
\hline
\end{tabular}
\egroup 
\; , \quad
\\
R^Z(\alpha|{w}) =& \frac{1}{2} \;
\bgroup 
\def\arraystretch{1.7} 
\begin{tabular}{|c|c|}
\hline
$\;\; 1 \;\;$ & $\;\; 0 \;\;$ \\
\hline
$1$ & $\;\; 0 \;\;$   \\
\hline
\end{tabular} \;.
\egroup
\end{split}
\end{equation}	
(The bottom left box of such a phase space diagram corresponds to phase space point $\alpha = (0,0)$ and the appropriate index increases by 1 when moving either up or right.) It corresponds to the spin up state in the $z$-direction for a qubit.
Form the instance of Eq.~\eqref{respect1} that results from this preparation and the identity channel:
\begin{equation}
\Delta R^Z(\beta|w)=\sum_{\{(S,B)|SB=Z\}}\Delta \left[ \sum_\alpha R_{ I}^S(\beta|\alpha)R^B(\alpha|w)\right].
\end{equation}
Applying the normalization rule
$\sum_\alpha R^S_I(\beta|\alpha) = 1$
to the terms with $B=X$ and $B=Y$, for which $R^B$
is uniform, yields a null contribution. What remains is
\begin{equation}
\Delta R^Z(\beta|w) = \Delta\left[ \sum_\alpha R_{I}^\mathcal{I}(\beta|\alpha)R^Z(\alpha|w)\right].
\end{equation}
For this to hold true, the transitions of $R_I^\mathcal{I}(\beta|\alpha)$ must not 
take either of the points on the nonzero
line of $R^Z(\alpha|w)$ away from that line. But this same argument can be made for a preparation corresponding to any other line and the form of $R_I^\mathcal{I}(\beta|\alpha)$ should not change. This implies 
\begin{equation} \label{identI}
R_I^\mathcal{I}(\beta|\alpha) = \delta_{\alpha\beta}.
\end{equation}
Therefore, whenever the identity transformation is inserted into Eq. \eqref{respect1},
the LHS of the equation will always equal the term on the RHS that includes $R_I^\mathcal{I}(\beta|\alpha)$. The other two terms---corresponding to the
symplectic matrices ${\mathcal R}$ and
${\mathcal L}$---must sum to zero, which can only be possible for all preparations when
\begin{equation} \label{identRL}
R_I^\mathcal{R}(\beta|\alpha) = R_I^\mathcal{L}(\beta|\alpha) = \frac{1}{4}.
\end{equation}
Eqs.~(\ref{identI}) and (\ref{identRL}) thus
give us the representation of the identity
transformation.

We now make use of Eq. \eqref{respect2} specialized to the case of a transformation being combined with its inverse.
Leveraging assumption B, 
we find
\begin{align} \label{UinverseU}
\Delta R^S_I(\gamma|\alpha) 
=\sum_{S'} \Delta\bigg[ \sum_\beta R_\mathcal{E}^{S S'}(\gamma|\beta) R^{S'}_{\mathcal{E}}(\alpha|\beta)\bigg].
\end{align}
We again use the fact that the sum
of $R_{\mathcal E}$ over its second
argument is unity.  From this it follows
that we can move the $\Delta$ on the 
right-hand side of
Eq.~(\ref{UinverseU}) to the factors inside the sum over $\beta$ (see Appendix B of Ref.~\cite{Braasch2021}):
\begin{equation} \label{Deltainside}
\begin{split}
\sum_{S'} \Delta\bigg[ \sum_\beta R_\mathcal{E}^{S S'}&(\gamma|\beta) R^{S'}_{\mathcal{E}}(\alpha|\beta)\bigg]
=\\
&\sum_{S',\beta} \Delta R_\mathcal{E}^{S S'}(\gamma|\beta) \Delta R^{S'}_{\mathcal{E}}(\alpha|\beta).
\end{split}
\end{equation}
And from Eq.~(\ref{RtoQ}), we have that
\begin{equation} \label{QDeltaR}
Q_{\mathcal E}(\gamma|\beta)
= \frac{1}{4} + \sum_S \Delta R_{\mathcal E}^S(\gamma|\beta).
\end{equation}
Combining Eqs.~(\ref{UinverseU}--\ref{QDeltaR}) with the inverse rule and the form of $R_I^\mathcal{I}(\beta|\alpha)$, one can show that the transition quasiprobabilities for any invertible transformation can be thought of as an orthogonal matrix:
\begin{align}
\sum_\beta & Q_\mathcal{E}(\gamma|\beta)  Q_\mathcal{E}(\alpha|\beta)  \nonumber \\
&= \sum_\beta \left(\frac{1}{4} + \sum_S\Delta R_\mathcal{E}^S(\gamma|\beta)\right) \left(\frac{1}{4} + \sum_{S'} \Delta R_\mathcal{E}^{S'}(\alpha|\beta)\right)\\
&= \frac{1}{4} + \sum_{S,S'} \Delta \left[ \sum_\beta R^{SS'}_\mathcal{E}(\gamma|\beta) R_\mathcal{E}^{S'}(\alpha|\beta)\right]\\
&= \frac{1}{4} +\sum_S \Delta R^S_I(\gamma|\alpha) \\
&= \delta_{\alpha\gamma}.
\end{align}

Although we ultimately want to know what sets
of transition probabilities $R_\mathcal{E}^S$ are allowed in our theory, the argument is less cumbersome if we work with the quasiprobabilities $Q_\mathcal{E}$ temporarily.
They are of course well defined in terms of the transition probabilities $R_\mathcal{E}^S$ (by \hbox{Eq.~(\ref{RtoQ}))}.

We can express the Wigner function 
$Q(\alpha|w)$ for a qubit as a four-component column vector $\vec{Q}_w$ on which a transition quasiprobability matrix 
$Q_{\mathcal E}$ acts.
Thus Eq.~(\ref{Wignertrans}) becomes
\begin{equation}
\vec{Q}_{w'} = Q_\mathcal{E} \vec{Q}_{w}\; ,
\end{equation}
where $w' = \mathcal{E}(w)$.
We also know that $Q_\mathcal{E}$ and its inverse both preserve normalization, which means that each row and each column of $Q_\mathcal{E}$ sums to unity.  
Because of this, we can write
\begin{equation}
\Delta \vec{Q}_{w'} = Q_\mathcal{E} \Delta \vec{Q}_{w}\; ,
\end{equation}
where $\Delta\vec{Q}$ is defined through our usual $\Delta$ notation, that is, by subtracting 1/4 from each component.
Now define the following orthogonal matrix:
\begin{equation}
M = \tfrac{1}{2}\mtx{rrrr}{1 & 1 & -1 & -1 \\ 1 & -1 & -1& 1 \\ 1 & -1 & 1 & -1 \\ 1 & 1 & 1 & 1}.
\end{equation}
Then we have
\begin{equation}
\left( M \Delta \vec{Q}_{w'}  \right) = M Q_\mathcal{E}  M^T \left( M \Delta \vec{Q}_{w} \right).
\end{equation}
Note that the last component of either $M\Delta \vec{Q}_{w}$ or $M\Delta\vec{Q}_{w'}$ is zero due to normalization.
So these vectors are confined to three
dimensions.  Meanwhile, the matrix $M Q_\mathcal{E} M^T$ is still orthogonal.  Moreover, it is block diagonal, consisting of a $3 \times 3$ block
in the upper left and the number 1 in the lower right.  Consequently, it is effectively a $3 \times 3$ orthogonal matrix acting on the three-dimensional
space in which  $M\Delta\vec{Q}_{w}$ can have nonzero 
components.  Let us
define $\hat{w}$ to be $\sum_\alpha Q(\alpha|w) \hat{A}_\alpha$. 
This matrix has unit trace, so we can express it as
$\hat{w} = (1/2)(I + \vec{r}\cdot \hat{\vec{\sigma}})$ for some real
vector $\vec{r}$. Then one can show from 
the definition of $\hat{A}_\alpha$ that the three nonzero components of $M\Delta\vec{Q}_{w}$
are the components $r_1$, $r_2$, $r_3$ of $\vec{r}$.  
Thus, the fact that $Q_\mathcal{E}$ is orthogonal implies that every reversible transformation can be thought of as a rotation of the Bloch sphere, possibly combined with
a reflection.

We now have a set of invertible transformations that is more permissive than that of a qubit, since
it includes the possibility of reflection.
That is, it includes transformations
represented by
$3 \times 3$ 
orthogonal matrices with determinant
$-1$.  (Or, in standard quantum terms,
it includes antiunitary
transformations.)  
However, not all of the negative-determinant
transformations are allowed, as we now
show.

One can always decompose a negative-determinant
orthogonal transformation of the sphere into an inversion through the center followed by a rotation.
We will denote the inversion operation as $\Omega$. To find $R_\Omega^S$, note that
the phase point operators are defined using an expansion of Pauli operators and have unit trace, so they can be represented by points in the same three-dimensional space in which $\vec{r}$ lives.  For example, $\hat{A}_{(0,0)}
= \tfrac{1}{2}(I + \vec{r}\cdot \hat{\vec{\sigma}})$, where $\vec{r} = (1,1,1)$, and more generally, we can 
write $\hat{A}_\alpha = \tfrac{1}{2}(I + \vec{r}_\alpha \cdot \hat{\vec{\sigma}})$.
Therefore, the inversion operation $\Omega(\hat{A}_\alpha)$ is well defined: extra minus signs appear before the $\hat{X}$, $\hat{Y}$, and $\hat{Z}$ terms.
We can then use Eqs. \eqref{Qdef} and \eqref{QtoR} to find
\begin{equation}
\begin{split}
&R^{\mathcal I}_\Omega(\beta | \alpha) = \tfrac{1}{2} - \delta_{\alpha \beta}, \\
&R^{\mathcal R}_\Omega(\beta | \alpha) = R^{\mathcal L}_\Omega(\beta | \alpha) = \tfrac{1}{4}.
\end{split}
\end{equation}
$R^\mathcal{I}_\Omega$ has negative values and therefore is not compatible with our formalism.

This does not yet rule out any of the other negative-determinant transformations. (It
does, however, rule out the possibility of
including even a single such transformation if
all the rotations are allowed, since we
could then construct $\Omega$.)
Again, all negative-determinant transformations can be written as an inversion followed by a rotation $\mathcal{E}$, 
and we can use the combination rule in Eq.~\eqref{respect2} to show that
\begin{equation} \label{eq:R-U-Omega}
R^S_{\mathcal{E} \circ \Omega}(\beta | \alpha) = \tfrac{1}{2} - R^S_\mathcal{E}(\beta | \alpha).
\end{equation}
It follows that we can only allow transformations described by $\mathcal{E}\circ \Omega$, if the rotation $\mathcal{E}$ is represented by transition probabilities that never exceed $1/2$.
In Appendix A, we show that this leaves us with only twelve possible rotations that can be composed with inversion to give legal transition probabilities. Let us call them
$\mathcal{E}_j$.
These include 90 degree righthand rotations around each of the six cardinal directions and 180 degree rotations around each axis that forms a 45 degree angle with a pair of cardinal axes. 

The twelve 
operations $\mathcal{E}_j \circ \Omega$
effect permutations of the four vectors $\vec{r}_\alpha$.  (Recall that
these vectors correspond to the four phase point operators and thus to the four points of phase space.) Composing these
operations, we get a set of twelve rotations of the form $\mathcal{E}_j\circ{\mathcal{E}}_k$. Altogether, this gives us a set of positive and negative-determinant transformations that correspond to the twenty-four ways one can permute the four phase point operators.
Although this set of twenty-four is quite
different from the set of
reversible transformations of a qubit, it is intriguing that it is a nontrivial set of transformations that can easily be understood classically.

We thus have two possibilities for the set of transformations: (i) a continuous set 
consisting of 
all the rotations of the Bloch sphere---the set we were aiming for---or (ii) a finite set that can be understood 
as comprising all possible permutations of the four ontic states.
Both sets are maximal in the sense that they cannot be augmented with any other transformations.

To summarize, it appears that our composition rules and auxiliary assumptions do not uniquely lead to qubit physics. 
Nonetheless, our simple setup does bring us remarkably close.
At this point, the best we can do is to include another assumption such as continuity of the set of transformations that would eliminate the finite set. 


\section{Conclusions} \label{conclusions}
For a quantum system with prime Hilbert-space dimension,
we have a way of decomposing the quantum description
of an experiment into a set of classical, epistemically
restricted descriptions.  For each of these classical
descriptions, which consist of nothing but probability
functions, we can imagine an observer using these
functions to
compute the probability of any given outcome of the 
experiment.  For any given classical observer, this
prediction will be a bad prediction, but we know how
to combine the predictions of all the classical
observers to recover the correct quantum mechanical
probability: we sum the nonrandom parts.  

But this picture begins with the standard formulation
of quantum theory.  Our aim is to develop an alternative,
self-contained
formulation of quantum theory in which the classical
descriptions are the primary mathematical entities. 
The formulation we seek would thus be based entirely on 
actual probability functions defined
on phase space.  
To create such a formulation, we need a set of criteria for determining
when a given probabilistic description of a preparation,
a transformation, or a measurement outcome is legitimate.
In this paper we have presented and begun to explore
a set of equations that might serve as the basis for
such criteria.  These equations---our
four composition rules---can all be placed
under the heading, ``sum the nonrandom parts.'' 
We have been led to this approach by the fact that
this intriguing prescription is the only
non-classical element of our formalism.  We are
wondering whether summing the nonrandom parts is
a key to what is characteristically ``quantum'' about
quantum theory, and we have speculated
that it may play a {fundamental} role {loosely} analogous to that
of the superposition principle in the
standard formulation.

Summing the nonrandom parts is a strange way to 
combine probability distributions.  It could easily
lead to illegal probabilities if there were not
some constraints on the probability distributions
being combined.  So simply by insisting that the
probabilities computed via this prescription are
legitimate, we are implicitly placing constraints on our
classical probability distributions.  This fact
has led us to ask the question: are those constraints, along with a 
set of intuitively plausible auxiliary 
assumptions,
sufficient to define the structure of quantum
theory?  

We have addressed this question for the case of a single
qubit, with only reversible transformations, and we have
found that we can recover the
usual quantum rules that determine what states are
allowed and what transformations are allowed.
(For the case of transformations, we need an 
assumption such as continuity to rule out a
particular finite set of unitary and antiunitary
transformations that is consistent with
our other assumptions.)

But the case of a single qubit is relatively simple.  
In our formalism, the set of allowed states is determined
entirely by the condition that
\begin{equation}
\sum_{B,\alpha} R^B(\alpha | w)^2 \le 1.
\end{equation}
For a general qudit, we have an analogous 
equation:
\begin{equation}
\sum_{B,\alpha} R^B(\alpha | w)^2 \le \frac{2}{d}.
\end{equation}
But for $d>2$, this is not the {\em only} condition
required for a state to be legitimate.  So any argument
from our composition rules is not
likely to be as simple as the one we were able to use for
a single qubit.  

Once we permit ourselves the extra assumption that the set of transformations is continuous, the reversible transformations on a single qubit are
also relatively simple.  They are equivalent to the
rotations in three dimensions.  So we mainly
needed to show that the matrix of 
transition quasiprobabilities, $Q_\mathcal{E}(\beta | \alpha)$,
is an orthogonal matrix.  In higher dimensions, this
matrix is again orthogonal, but other conditions must
also be met in order to arrive at the unitary 
transformations.  

However, we have by no means used all the information 
available in our composition rules [Eqs.~(\ref{respect1}--\ref{respect4})].
So one can hope that these equations constitute
a sufficient or nearly sufficient set of restrictions for arbitrary prime $d$.

{Ultimately we would like to extend our work
to all Hilbert-space dimensions.}  
In the analysis of Ref.~\cite{Wootters1987}, a system with composite
dimension $d$ is treated as a composite system.
It would be natural for us to use the same strategy
here.  Thus, the phase space for a system with dimension
6 would be a four-dimensional space, the 
Cartesian product of ${\mathbb Z}_2^2$ and
${\mathbb Z}_3^2$.  A framework for a preparation
or a measurement outcome would consist of a striation
$B_2$ of ${\mathbb Z}_2^2$ and a striation 
$B_3$ of ${\mathbb Z}_3^2$.  {In future
work, we plan to use this factorization scheme
to extend our treatment of 
preparations and measurements to arbitrary
composite dimensions, and indeed, to arbitrary
composite discrete systems.  We see no obstacles
there.  The treatment of 
{\em transformations}, on the other hand, is more challenging.
One can show that, if $S$ ranges over all the
$2 \times 2$ symplectic matrices with entries
in ${\mathbb Z}_d$, where $d$ is composite, then
the whole set of probability distributions
$R^S_\mathcal{E}(\beta | \alpha)$, defined in the natural
way, does not contain the information needed
to reconstruct the transition quasiprobabilities
$Q_\mathcal{E}(\beta | \alpha)$.  (Moreover,
factoring the group
of symplectic transformations into groups 
associated with the prime factors of $d$ 
does not change the information content
of the $R$'s.)
This fact does not imply that 
our formalism cannot be extended to composite
dimensions, but it does
mean that new ideas will be needed.}

{Of course we would also like to
extend our ``classical'' treatment of quantum 
theory to the case of continuous phase space, the
realm that is truly the domain of classical 
mechanics.  The concepts of striations, displacements,
and symplectic transformations are all sensible 
concepts for such a phase space.  But we anticipate
challenges in finding the proper analogue of our 
notion of the sum of the nonrandom parts.  For example,
whereas the ``random part'' of
a probability distribution over a discrete phase space
is simply the uniform distibution, there is no 
such thing as a normalized uniform distribution
over an infinite phase space. We plan to address
this and related issues in future work.}

Finally, it is interesting to ask whether our formalism
lends itself to an ontological account of a quantum 
experiment. Each of our imagined classical observers would
have no problem finding a realistic interpretation of 
their description of the experiment: the system is always
at some location in phase space, and when a transformation
occurs, the system jumps probabilistically to some other point.
It is much more difficult, though, to find an ontological
account that incorporates all the classical descriptions. 
This is not to say it cannot be done.  If it is 
possible, it will 
certainly require expanding the picture beyond that
of a stochastic process on phase space.  And it would
require making physical sense of the mathematical
prescription to sum the nonrandom parts.

\section*{Appendix}

Here we prove that there is one possible set of transformations allowed within our scheme that includes a finite set of negative-determinant orthogonal transformations of the Bloch sphere.

Recall that we started with only our set of rules and assumptions and made no reference to Hilbert space.
Using these, we found that the legal transformations can be understood as 
orthogonal transformations of the Bloch sphere.
We now wonder what the possible rotations are that can be composed with the inversion operations $\Omega$ without generating
a negative probability.  We have seen
that such a rotation must itself never generate
a probability greater than 1/2 for any
of its $R$ values.  
We could compute the $R$ values for 
the orthogonal transformation $Q_{\mathcal E}$
directly from Eq.~(\ref{QtoR}). But
since we have established a correspondence between rotations of the sphere and unitary transformations, it is legitimate
to use Eq.~(\ref{QUdef}), together with
Eq.~(\ref{QtoR}), to compute 
these values. 

The virtue of this strategy is that
any unitary operation on a qubit can be expressed in a simple way: 
\begin{equation}  \label{Ufrtomw}
\hat{U} = u_0 \hat{I} + iu_1 \hat{X} + iu_2 \hat{Y} + iu_3 \hat{Z},
\end{equation}
where $\vec{u} = (u_0, u_1, u_2, u_3)$ is a real four-vector with unit length.  
The set of functions $R_{\mathcal E}^S$ (where $\mathcal{E}$ refers to this 
unitary transformation) holds twelve 
values that we can calculate using \hbox{Eqs.~(\ref{QUdef}, \ref{QtoR})}.
(For each of the three $S$'s, $R_{\mathcal E}^S$ has sixteen entries, but remember that these are partitioned into four displacement classes, each of which
holds a single value of 
$R_{\mathcal E}^S$.)
These values are listed in the following phase space diagrams, where the label on the left is the symplectic transformation $S$ and the phase space points correspond to the value $\delta = \beta - S \alpha$.

\begin{equation}  \label{tablesw}
\begin{split}
&{\mathcal I}: \hspace{11mm} 
\bgroup 
\def\arraystretch{1.7} 
\begin{tabular}{|c|c|}
\hline
$u_3^2$ & $u_2^2$ \\
\hline
$u_0^2$ & $u_1^2$   \\
\hline
\end{tabular} 
\egroup \\
&{\mathcal R}: \hspace{3mm}\frac{1}{4}  \times  
\bgroup 
\def\arraystretch{1.7} 
\begin{tabular}{|c|c|}
\hline
$(u_0 - u_1 + u_2 - u_3)^2$ & $(u_0 + u_1 - u_2 - u_3)^2$ \\
\hline
$(u_0 + u_1 + u_2 + u_3)^2$ & $(u_0 - u_1 - u_2 +u_3)^2$ \\
\hline
\end{tabular} 
\egroup \\
&{\mathcal L}: \hspace{4mm}\frac{1}{4}  \times  
\bgroup 
\def\arraystretch{1.7} 
\begin{tabular}{|c|c|}
\hline
$(u_0 - u_1 + u_2 + u_3)^2$ & $(u_0 + u_1 + u_2 - u_3)^2$ \\
\hline
$(u_0 - u_1 - u_2 - u_3)^2$ & $(u_0 + u_1 - u_2 +u_3)^2$ \\
\hline
\end{tabular} 
\egroup \\
\end{split}
\end{equation}

Suppose for now that all the $u_j$'s are nonnegative.  Again, we have already seen in the main text that in order to be composed with the inversion operator $\Omega$, a transformation can have no value of $R^S_\mathcal{E}(\beta|\alpha)$ greater than 1/2.
Therefore, we must have the following three relations:
\begin{equation}
\begin{split}
&u_0^2 + u_1^2 + u_2^2 + u_3^2 = 1, \\
&u_j \le \frac{1}{\sqrt{2}} , \\
&u_0 + u_1 + u_2 + u_3 \le \sqrt{2},
\end{split}
\end{equation}
where the second line is from $R^{\mathcal{I}}_{\mathcal E}$ and the third line is from $R^{\mathcal{R}}_{\mathcal E}$.
The argument is easier to see if we define $v_j = \sqrt{2} \, u_j$.  Then the conditions on $v_j$ are
\begin{equation}
\begin{split}
&v_0^2 + v_1^2 + v_2^2 + v_3^2 = 2, \\
&v_j \le 1, \\  
&v_0 + v_1 + v_2 + v_3 \le 2.
\end{split}
\end{equation}
Because each $v_j$ is no larger than 1, $v_j^2 \le v_j$. But then the only way to satisfy the first and third conditions is to make each $v_j^2$ {\em equal} to $v_j$.  This means
each $v_j$ must be either 0 or 1.  And then the first equation tells us that exactly two $v_j$'s are equal to 1 and the other two are equal to 0.  So, exactly two of the $u_j$'s are
equal to $1/\sqrt{2}$ and the other two are equal to zero.  

Of course we also have to deal with the possibility that one or more of the $u_j$'s is negative.  But looking at the values of $R$ in Eq.~(\ref{tablesw}), we see that all
possible combinations of plus and minus signs appear in $R^{\mathcal R}_\mathcal{E}$ and $R^{\mathcal L}_\mathcal{E}$.  So we can replace the last $v$ condition with 
$|v_0| + |v_1| + |v_2| + |v_3| \le 2$.  And the middle equation can be $|v_j| \le 1$.  Then the same argument applies.

The only option we are left with is to form four-vectors where only two entries are $\pm 1/\sqrt{2}$ and the other two are zero. There are twenty-four ways to do this, but one half of that set of vectors is just the negative of the other half. Mapping back from $\vec{u}$ to $\hat{U}$, this minus sign is just a phase that can be ignored. We are left with twelve possible rotations compatible with the inversion operator.

\end{document}